\documentclass[11pt,eqs]{article}
\usepackage{latexsym}
\usepackage{amsmath}
\usepackage{graphicx,relsize}
\usepackage{amstext,amssymb}
\usepackage{amsmath,tikz}

\def\cleq{\setcounter{equation}{0}}
\title{
Effective theories of two T-dual theories are also T-dual
\thanks{Work supported in part by
the Serbian Ministry of Education and Science, under contract No. 171031.}}
\author{Lj. Davidovi\'c \thanks{e-mail: ljubica@ipb.ac.rs} and B. Sazdovi\'c
\thanks{e-mail: sazdovic@ipb.ac.rs}\\
{\it Institute of Physics,}\\
{\it University of Belgrade,}\\
{\it 11001 Belgrade, P.O.Box 57, Serbia}}

\begin{document}
\maketitle
\begin{abstract}
We investigate 
how T-duality and solving the boundary conditions of the open bosonic string are related.
We start by considering the T-dualization of the open string  moving 
in the constant background.
 We take that the coordinates of the initial theory satisfy either Neumann or Dirichlet boundary conditions.
It follows that the coordinates of  T-dual theory satisfy exactly the opposite set of boundary conditions.
We treat the boundary conditions of both theories as constraints, and apply the Dirac procedure to them, which 
results in forming $\sigma$-dependent constraints.
We solve these constraints and obtain the effective theories for the solution.
We show that 
the effective closed string theories 
are also T-dual.
\end{abstract}

\section{Introduction}

T-duality \cite{BGS,KY,SS},  first observed in string theory,  interchanges the string momenta and winding numbers, leaving the spectrum unchanged. Its description on the string sigma model level was first given by Buscher \cite{B1,B2}.
The Buscher procedure \cite{RV} covers the T-dualization
of the coordinates on which the background fields do not depend.
The generalized Buscher procedure, applicable to the arbitrary coordinate of the coordinate dependent background was proposed in \cite{DST}. The T-dual theory obtained by this prescription 
is nongeometric, described in terms of the dual coordinates and their double.
The double field theories are investigated in \cite{DFT1,DFT2}.

The nongeometricity appears naturally when considering the open bosonic string moving in a weakly curved background with all coordinates satisfying the
Neumann boundary conditions. The problem of solving these boundary conditions was considered in \cite{DS,DS1,DS2}.
 In the first two papers the conditions were treated as constraints in a Dirac procedure. 
In the third, the solution of boundary conditions was presupposed in a form expressing the odd coordinate and momenta parts in terms of their even parts. Both treatments lead to effective theories, obtained for the solution of boundary condition, defined in nongeometric space given in terms of even parts of  coordinates and of their doubles.

In this paper we consider the open string moving in the constant
background fields: metric $G_{\mu\nu}$ and antisymmetric Kalb-Ramond field
$B_{\mu\nu}$.
It is well known that the 
constant Kalb-Ramond field does not affect the 
dynamics in the world-sheet interior but it contributes  to its boundary and causes the noncommutativity of the string coordinates.
Also, we consider the T-dual theory, obtained applying the T-dualization procedure to the above theory. The T-dual theory has a standard action describing the T-dual string moving in the background with a T-dual metric $
{^\star G}^{\mu\nu}=
(G^{-1}_{E})^{\mu\nu}$, which is an inverse of the effective metric
and a dual Kalb-Ramond field
$
^{\star}B^{\mu\nu}=
\frac{\kappa}{2}\theta^{\mu\nu}$
 which is the noncommutativity parameter,
in Seiberg--Witten terminology of the open bosonic string
theory \cite{SW}.

We consider the mixed boundary conditions, for both initial and T-dual coordinates and solve them using techniques developed in \cite{DS,DS1,DS2}.
We chose the Neumann boundary conditions for coordinate directions $x^{a}$  and Dirichlet conditions for the rest of the coordinates $x^{i}$ of the initial theory.
As usual, using the T-dual coordinate transformation laws one shows
that the chosen boundary conditions of the initial theory transform to the boundary conditions of the T-dual theory, so that $y_{a}$ satisfy the Dirichlet and $y_{i}$ the Neumann boundary conditions.

We treat all boundary conditions as constraints and follow the Dirac procedure.
The new constraints are found, first as a Poisson bracket between 
the hamiltonian and
the  boundary conditions, and every subsequent as a Poisson bracket between 
the hamiltonian and 
the previous constraint. Using the Taylor expansion,
we represent this infinite set of constraints we obtain, 
by only two $\sigma$-dependent constraints \cite{SN,SN1,SN2,SN3}, one for each endpoint.
Imposing $2\pi-$periodicity, to the variables building the constraints, one observes that
the constraints at $\sigma=\pi$ can be expressed
in terms of that at $\sigma=0$,
and that in fact solving one pair of constraints one solves the other pair as well.

We can separate the constraints into 
even and odd parts under world-sheet parity transformation
($\Omega : \sigma \rightarrow -\sigma$),
separating the variables building the constraints into even and odd parts.
Solving the $\sigma$-dependent constraints, one reduces the phase space by half.
Halves of the original canonical variables
are treated as effective variables:
the independent variables and their canonical conjugates.
For the solution of the constraints we obtain the effective theories,
defined in terms of the effective variables. We examine their characteristics and
confirm that the effective theories of two T-dual theories are also T-dual.

The paper is organized as follows:
In section \ref{sec:ost} we consider the standard open bosonic string action
and we choose the boundary conditions for every coordinate.
Then, we find the T-dual theory,
and show that T-dual coordinates satisfy exactly the opposite boundary condition for a given direction of the T-dual space-time, than for the corresponding direction of the original space-time.
In section \ref{sec:cbc}, we rewrite the boundary conditions in the canonical form
and find the new constraints following the Dirac procedure. We gather the constraints  into $\sigma-$dependent constraints,  separate the canonical variables
into their even and odd parts, and solve  the constraints.
In section \ref{sec:nev} we find the noncommutativity relations for coordinates and momenta of both initial and T-dual theories.
In section \ref{sec:eff} we calculate the effective theories,
which will be obtained from the initial theories for the solution of the constraints.
We show that the effective theories of the initial and T-dual theory remain T-dual,
and find the effective T-duality coordinate transformation laws.

\section{The open bosonic string and its T-dual}\label{sec:ost}

The bosonic string sigma model, describes the bosonic string moving in a curved background 
associated with the massless bosonic fields: a metric field $G_{\mu\nu}$, a Kalb-Ramond field $B_{\mu\nu}$ and a dilaton field $\Phi$.
The dynamics is described by the action \cite{BBS,Z}
\begin{eqnarray}\label{eq:action0}
S[x]&=& \kappa \int_{\Sigma} d^2\xi\sqrt{-g} \Big[\Big(\frac{1}{2}{g}^{\alpha\beta}G_{\mu\nu}(x)
+\frac{\varepsilon^{\alpha\beta}}{\sqrt{-g}}B_{\mu\nu}(x)\Big)
\partial_{\alpha}x^{\mu}\partial_{\beta}x^{\nu}+\Phi(x)R^{(2)}\Big].
\end{eqnarray}
The integration goes over two-dimensional world-sheet $\Sigma$
parametrized by
$\xi^\alpha$ ($\xi^{0}=\tau,\ \xi^{1}=\sigma$),
$g_{\alpha\beta}$ is the intrinsic world-sheet metric, $R^{(2)}$ corresponding 2-di\-me\-nsi\-o\-nal scalar curvature,
$x^{\mu}(\xi),\ \mu=0,1,...,D-1$ are the coordinates of the
D-dimensional space-time,
$\kappa=\frac{1}{2\pi\alpha^\prime}$
with $\alpha^\prime$ being the Regge slope parameter
and $\varepsilon^{01}=-1$.
The space-time fields in which the string moves have to obey the space-time equations of motion, in order to have a conformal invariance on the quantum level. 
If the dilaton field is taken to be zero,
and the conformal gauge is considered $g_{\alpha\beta}=e^{F}\eta_{\alpha\beta}$, the action can be rewritten as
\begin{equation}\label{eq:action0}
S=\kappa\int d\xi^{2}\partial_{+}x^\mu\Pi_{+\mu\nu}\partial_{-}x^\nu,
\end{equation}
with the background field composition
\begin{equation}\label{eq:pi}
\Pi_{\pm\mu\nu}(x)=
B_{\mu\nu}(x)\pm\frac{1}{2}G_{\mu\nu}(x),
\end{equation}
and the light-cone coordinates given by
\begin{equation}
\xi^{\pm}=\frac{1}{2}(\tau\pm\sigma),
\qquad
\partial_{\pm}=
\partial_{\tau}\pm\partial_{\sigma}.
\end{equation}

From the minimal action principle one obtains the equations of motion and the boundary conditions 
\begin{equation}\label{eq:bcdef}
\gamma^{(0)}_{\mu}\delta x^\mu\Big{|}_{0}^{\pi}=0,
\end{equation}
where
\begin{equation}
\gamma^{(0)}_\mu=\kappa\big(
\Pi_{+\mu\nu}\partial_{-}x^\nu
+\Pi_{-\mu\nu}\partial_{+}x^\nu\big).
\end{equation}

For the closed string the boundary conditions are fulfilled because of the periodicity of its coordinates.
In the open string case, for each of the space-time coordinates one can fulfill the boundary conditions (\ref{eq:bcdef})
by choosing either
the Neumann
or the Dirichlet boundary condition.
Let us choose the Neumann condition for coordinates $x^{a},\,a=0,1,\dots,p$ and the Dirichlet condition for coordinates $x^{i},\,i=p+1,\dots,D-1$,
which read
\begin{eqnarray}\label{eq:bcnd}
&& {\it Neumann}:\quad
\gamma^{(0)}_{a}\Big{|}_{\partial\Sigma}=0,\quad
{_{\mathsmaller N}\gamma}^{0}_{a}
\equiv
\gamma^{(0)}_{a}=
\kappa\big(
\Pi_{+ab}\partial_{-}x^b
+\Pi_{-ab}\partial_{+}x^b\big),
\nonumber\\
&&{\it Dirichlet}:\quad
\kappa\dot{x}^{i}\Big{|}_{\partial\Sigma}=0,\quad{_{\mathsmaller D}}\gamma^{i}_{0}\equiv\kappa\dot{x}^{i}.
\end{eqnarray}
We consider the block diagonal constant 
metric and Kalb-Ramond field $G_{\mu\nu}=const$, $B_{\mu\nu}=const$
\begin{equation}\label{eq:bg}
G_{\mu\nu}=\left(\begin{array}{cc}
G_{ab} & 0\\
0 & G_{ij}         
\end{array}\right),\quad
B_{\mu\nu}=\left(\begin{array}{cc}
B_{ab} & 0\\
0 & B_{ij}         
\end{array}\right).
\end{equation}

\subsection{Open string theory T-dual}

Let us find a T-dual of the open string theory described by the 
action (\ref{eq:action0}).
In order to find the T-dual action, one substitutes the ordinary derivatives 
with the covariant derivatives $D_\pm x^\mu=\partial_\pm x^\mu+v_\pm^\mu$,
defined in terms of the gauge fields $v^\mu_\pm$.
One adds the Lagrange multiplier term to make the introduced gauge fields unphysical.
The gauge is fixed taking $x^\mu(\xi)=0$.
Next, one finds the equations of motion varying the obtained gauge fixed action over the gauge fields $v^\mu_\pm$.
The T-dual action is obtained by substituting the expressions
for the gauge fields obtained from these equations of motion, into the gauge fixed action.
The T-dual action reads \cite{DS}
\begin{equation}\label{eq:tdual}
^\star S=
\frac{\kappa^{2}}{2}\int d\xi^{2}\partial_{+}y_\mu
\Theta_{-}^{\mu\nu}\partial_{-}y_\nu,
\end{equation}
The dual background field composition equals
\begin{eqnarray}\label{eq:dpolja}
^\star\Pi^{\mu\nu}_\pm=\frac{\kappa}{2}\Theta^{\mu\nu}_\mp=-(G^{-1}_{E}\Pi_\mp G^{-1})^{\mu\nu},
\quad (G_{E})_{\mu\nu}=(G-4BG^{-1}B)_{\mu\nu},
\end{eqnarray}
where $G_{E}$ is the effective metric.
The T-dual metric is its inverse 
\begin{equation}
{^\star G}^{\mu\nu}=
(G^{-1}_{E})^{\mu\nu},
\end{equation}
and a T-dual Kalb-Ramond field is
\begin{equation}
^{\star}B^{\mu\nu}=
\frac{\kappa}{2}\theta^{\mu\nu},
\end{equation}
where $\theta^{\mu\nu}=-\frac{2}{\kappa}(G^{-1}_{E}BG^{-1})^{\mu\nu}$ is the noncommutativity parameter.

Because of the choice (\ref{eq:bg}), the composition of the T-dual background fields is also block diagonal
\begin{equation}\label{eq:teta}
\Theta^{\mu\nu}_\pm=\left(\begin{array}{cc}
\Theta^{ab}_\pm & 0\\
0 & \Theta^{ij}_\pm         
\end{array}\right),
\end{equation}
given in terms of the inverse of the initial metric and the effective metric
\begin{eqnarray}\label{eq:metinv}
(G^{-1})^{\mu\nu}=\left(\begin{array}{cc}
({G}^{-1})^{ab} & 0\\
0 & ({G}^{-1})^{ij}       
\end{array}\right),\quad
(G_{E})_{\mu\nu}=\left(\begin{array}{cc}
({G}_{E})_{ab} & 0\\
0 & ({G}_{E})_{ij}         
\end{array}\right),
\end{eqnarray}
by
\begin{eqnarray}\label{eq:nulte}
&&\Theta^{ab}_{\pm}=
-\frac{2}{\kappa}(G_{E}^{-1})^{ac}\,\Pi_{\pm cd}({{G}}^{-1})^{db}
={\theta}^{ab}\mp \frac{1}{\kappa}({G}_{E}^{-1})^{ab},
\nonumber\\
&&\Theta^{ij}_{\pm}=
-\frac{2}{\kappa}(G_{E}^{-1})^{ik}\,\Pi_{\pm kl}({{G}}^{-1})^{lj}
={\theta}^{ij}\mp \frac{1}{\kappa}(G_{E}^{-1})^{ij},
\end{eqnarray}
where
$(G_{E})_{ab}=G_{ab}-4B_{ac}({G}^{-1})^{cd}B_{db}$
and $(G_{E})_{ij}=G_{ij}-4B_{ik}({G}^{-1})^{kl}B_{lj}$.
The components of the non-commutativity parameter are
\begin{eqnarray}
&& \theta^{ab}=
-\frac{2}{\kappa}(G_{E}^{-1})^{ac}\,B_{cd}({{G}}^{-1})^{db},
\nonumber\\
&& {\theta}^{ij}=
-\frac{2}{\kappa}(G_{E}^{-1})^{ik}\,B_{kl}({{G}}^{-1})^{lj}.
\end{eqnarray}

The coordinates of the initial and the T-dual theory are connected by T-duality coordinate transformation laws,
which read
\begin{eqnarray}\label{eq:rel}
&& \partial_{\pm}x^\mu\cong
-\kappa\Theta^{\mu\nu}_{\pm}
\partial_{\pm} y_\nu,
\nonumber\\
&&
\partial_{\pm}y_\mu\cong
-2\Pi_{\mp\mu\nu}\partial_{\pm}x^\nu.
\end{eqnarray}
The T-dual boundary conditions are
\begin{equation}\label{eq:bcdefdual}
^\star\gamma^{(0)\mu}\,\delta y_\mu\Big{|}_{0}^{\pi}=0,
\end{equation}
where
\begin{equation}
{^\star\gamma^{(0)\mu}}=\frac{\kappa^{2}}{2}\Big[
\Theta_{-}^{\mu\nu}\partial_{-}y_\nu
+\Theta_{+}^{\mu\nu}\partial_{+}y_\nu
\Big].
\end{equation}
The T-dual theory (\ref{eq:tdual})
is equivalent to a 
open string theory
(\ref{eq:action0}) with chosen boundary conditions (\ref{eq:bcnd}),
if the T-dual boundary conditions are fulfilled 
in a Neumann way for coordinates $y_i$ and in a Dirichlet way for $y_a$
\begin{eqnarray}\label{eq:dualbc}
&&{\it Neumann}:\quad ^\star\gamma^{(0)i}\Big{|}_{\partial\Sigma}=0,\qquad_{\mathsmaller N}^\star\gamma^{i}_{0}\equiv\,^\star\gamma^{(0)i}=\frac{\kappa^{2}}{2}\Big[
\Theta_{-}^{ij}\partial_{-}y_j
+\Theta_{+}^{ij}\partial_{+}y_j
\Big],
\nonumber\\
&&{\it Dirichlet}:\qquad\kappa\dot{y}_{a}\Big{|}_{\partial\Sigma}=0,\qquad_{\mathsmaller D}^\star\gamma_{a}^{0}\equiv\kappa\dot{y}_{a}.
\end{eqnarray}
This is because of the T-duality transformation law (\ref{eq:rel}), which gives
\begin{equation}
-\kappa\dot{x}^\mu\cong{^\star\gamma}^{(0)\mu}(y),
\quad
{\gamma}_\mu^{(0)}(x)\cong-\kappa\dot{y}_\mu,
\end{equation}
and consequently
\begin{eqnarray}\label{eq:bcrel}
&&{_{\mathsmaller D}}\gamma^{i}_{0}\equiv\kappa\dot{x}^{i}\cong
-{^\star\gamma}^{(0)i}\equiv\,-
^\star_{\mathsmaller N}\gamma^{i}_{0},
\nonumber\\
&&{_{\mathsmaller N}\gamma}^{0}_{a}\equiv \gamma^{(0)}_{a}
\cong-\kappa\dot{y}_{a}=-\,
_{\mathsmaller D}^\star\gamma_{a}^{0}.
\end{eqnarray}
So, performing T-dualization one  changes the type of the boundary conditions which the coordinates in $i$ and $a$ directions satisfy.


\section{Dirac consistency procedure applied to the boundary conditions}\label{sec:cbc}
\cleq

The coordinates of the initial 
and T-dual open string satisfy the appropriate set of the boundary conditions (\ref{eq:bcnd}) and (\ref{eq:dualbc}),
obtained from the  actions (\ref{eq:action0}) and  (\ref{eq:tdual}). 
In this section, we are going to treat them as constraints and we will apply the Dirac consistency procedure.
In order to implement the procedure,
let us find the canonical form of the boundary conditions, and express them in terms of the currents building the energy-momentum tensors, and consequently the hamiltonians.

The momenta conjugated to the coordinates of the initial and T-dual theories (\ref{eq:action0}) and  (\ref{eq:tdual}) are
\begin{eqnarray}\label{eq:momenta}
\pi_\mu&=&-2\kappa B_{\mu\nu}x^{\prime\nu}+\kappa G_{\mu\nu}\dot{x}^\nu,
\nonumber\\
^\star\pi^\mu&=&
-\kappa^{2} \theta^{\mu\nu}y^{\prime}_{\nu}+\kappa (G^{-1}_{E})^{\mu\nu}\dot{y}_\nu=
-2\kappa\, {^\star B}^{\mu\nu}y^{\prime}_{\nu}+\kappa\, {^\star G}^{\mu\nu}\dot{y}_\nu.
\end{eqnarray}

The energy-momentum tensor components for the initial theory
can be expressed in terms of currents
\begin{equation}\label{eq:struja}
j_{\pm\mu}=\pi_\mu+2\kappa\Pi_{\pm\mu\nu}x^{\prime\nu},
\end{equation}
as
\begin{equation}\label{eq:tpm}
T_\pm=\mp\frac{1}{4\kappa}(G^{-1})^{\mu\nu}j_{\pm\mu}j_{\pm\nu}.
\end{equation}
Using the first relation in (\ref{eq:momenta}), the currents can be rewritten in 
terms of coordinates as
\begin{equation}\label{eq:jk}
j_{\pm\mu}=\kappa G_{\mu\nu}\partial_\pm x^\nu.
\end{equation}
The canonical hamiltonian density is
\begin{equation}
{\cal H}_{C}=T_{-}-T_{+}=\frac{1}{4\kappa}(G^{-1})^{\mu\nu}\Big[
j_{+\mu}j_{+\nu}
+j_{-\mu}j_{-\nu}
\Big].
\end{equation}

The hamiltonian density and the energy-momentum tensor of the T-dual theory
\begin{eqnarray}
&&^\star T_\pm=\mp\frac{1}{4\kappa}(^\star G^{-1})^{\mu\nu}\,{^\star j}_{\pm\mu}{^\star j}_{\pm\nu},
\nonumber\\
&&^\star{\cal H}_{C}={^\star T}_{-}-{^\star T}_{+}=\frac{1}{4\kappa}(^\star G^{-1})^{\mu\nu}\Big[
{^\star j}_{+\mu}{^\star j}_{+\nu}
+{^\star j}_{-\mu}{^\star j}_{-\nu}
\Big]
\end{eqnarray}
are expressed in terms of the dual currents
given by 
\begin{equation}\label{eq:strujadual}
^\star j_{\pm}^{\mu}={^\star\pi}^\mu+2\kappa{^\star\Pi}_{\pm}^{\mu\nu}y^{\prime}_{\nu},
\end{equation}
where ${^\star\Pi}_{\pm}^{\mu\nu}$ is defined in (\ref{eq:dpolja}).
Using the second relation in (\ref{eq:momenta}) one obtains
\begin{equation}\label{eq:jkd}
^\star j_{\pm}^{\mu}=\kappa(G^{-1}_{E})^{\mu\nu}\partial_\pm y_\nu.
\end{equation}

\subsection{The Dirac procedure applied to the initial theory}

Let us treat 
the Neumann and Dirichlet boundary conditions (\ref{eq:bcnd}) of the initial theory 
as canonical constraints and apply the Dirac consistency procedure to them, following \cite{DS,DS1}.
The simplest way to obtain the
explicit form of these constraints is using the currents defined in (\ref{eq:struja}).
Because the hamiltonian is already expressed in terms of these currents, all that remains
is to find their algebra. 

The algebra of currents \cite{DS18} in a constant background is given by
\begin{eqnarray}
&&
\big{\{}j_{\pm\mu}(\sigma),j_{\pm\nu}(\bar\sigma)\big{\}}=
\pm\,2\kappa\, G_{\mu\nu}\,\delta^\prime(\sigma-\bar\sigma),
\nonumber\\
&&\big{\{}j_{\pm\mu}(\sigma),j_{\mp\nu}(\bar\sigma)\big{\}}=0.
\end{eqnarray}
Using the expressions for momenta (\ref{eq:momenta}), 
one can rewrite the Neumann (N) and Dirichlet (D) boundary conditions (\ref{eq:bcnd}) in a canonical form
\begin{eqnarray}\label{eq:bccan}
_{_{N}}\!\gamma^{0}_{a}&=&
\Pi_{+ab}(G^{-1})^{bc}j_{-c}
+\Pi_{-ab}(G^{-1})^{bc}j_{+c},
\nonumber\\
_{_{D}}\!\gamma^{i}_{0}&=&\kappa\dot{x}^{i}=\frac{1}{2}(G^{-1})^{ij}(j_{+j}+j_{-j}).
\end{eqnarray}
Following the Dirac procedure,
one can impose consistency to these constraints.
The additional constraints are defined for every $n\ge 1$ by
\begin{equation}
_{_{N}}\!\gamma^{n}_{a}=\{H_{C},{_{_{N}}}\!\gamma^{n-1}_{a}\},
\quad
_{_{D}}\!\gamma^{i}_{n}=\{H_{C},{_{_{D}}}\!\gamma^{i}_{n-1}\},
\end{equation}
with $H_{C}=\int\,d\sigma{\cal H}_{C}$ being the canonical hamiltonian.

All these constraints can be gathered into only two constraints, which depend on the space parameter of the world-sheet. We will 
multiply every constraint by an appropriate degree of the world-sheet space parameter $\sigma$
and add the terms together, forming
two sigma dependent constraints
\begin{equation}
\Gamma^{N}_{a}(\sigma)=\sum_{n\geq0}\frac{\sigma^{n}}{n!}\, 
{_{_{N}}}\!\gamma^{n}_{a}
\Big{|}_{\sigma=0},
\quad
\Gamma^{i}_{D}(\sigma)=\sum_{n\geq0}\frac{\sigma^{n}}{n!}\,
{_{_{D}}}\!\gamma_{n}^{i}\Big{|}_{\sigma=0}.
\end{equation}

Because the background fields are constant, 
the Poisson bracket
between the hamiltonian and the currents will produce the first $\sigma$-derivative of the 
currents
\begin{equation}
\{H_{C},j_{\pm\mu}(\sigma)\}=\mp j^\prime_{\pm\mu}(\sigma).
\end{equation}
 Consequently, the $n$-th constraints will be given in terms of 
the $n-$th derivative of the currents
\begin{eqnarray}\label{eq:conj}
&&_{_{N}}\!\gamma^{n}_{a}=
\Pi_{+ab}(G^{-1})^{bc}j^{(n)}_{-c}
+\Pi_{-ab}(G^{-1})^{bc}(-1)^{n}j_{+c}^{(n)},
\nonumber\\
&&_{_{D}}\!\gamma^{i}_{n}=\frac{(G^{-1})^{ij}}{2}\Big{[}(-1)^{n}j_{+j}^{(n)}+j_{-j}^{(n)}\Big{]}.
\end{eqnarray}
So, the constraints read
\begin{eqnarray}
&&\Gamma^{N}_{a}(\sigma)=\sum_{n\geq0}\frac{\sigma^{n}}{n!}
\Big[
\Pi_{+ab}(G^{-1})^{bc}j^{(n)}_{-c}
+\Pi_{-ab}(G^{-1})^{bc}(-1)^{n}j_{+c}^{(n)}
\Big]\Big{|}_{\sigma=0},
\nonumber\\
&&\Gamma^{i}_{D}(\sigma)=\frac{1}{2}\sum_{n\geq0}\frac{\sigma^{n}}{n!}
(G^{-1})^{ij}\Big[(-1)^{n}j_{+j}^{(n)}+j_{-j}^{(n)}\Big]\Big{|}_{\sigma=0},
\end{eqnarray}
where $(n)$ marks the n-th partial derivative over $\sigma$.
Summing, we obtain the explicit form of the sigma dependent constraints
\begin{eqnarray}\label{eq:gamma}
&&\Gamma^{N}_{a}(\sigma)=
\Pi_{+ab}(G^{-1})^{bc}j_{-c}(\sigma)
+\Pi_{-ab}(G^{-1})^{bc}
j_{+c}(-\sigma),
\nonumber\\
&&\Gamma^{i}_{D}(\sigma)=\frac{1}{2}
(G^{-1})^{ij}\Big[
j_{+j}(-\sigma)
+j_{-j}(\sigma)\Big].
\end{eqnarray}

The Poisson brackets of $\sigma$-dependent constraints are
\begin{eqnarray}
&&\{\Gamma^{N}_{a}(\sigma),\Gamma^{N}_{b}(\bar\sigma)\}=
-\kappa
(G_{E})_{ab}\,
\delta^\prime(\sigma-\bar\sigma),
\nonumber\\
&&\{\Gamma^{i}_{D}(\sigma),\Gamma^{j}_{D}(\bar\sigma)\}=-2\kappa(G^{-1})^{ij}\delta^\prime(\sigma-\bar\sigma).
\end{eqnarray}
Therefore, they are of the second class and one can solve them.

Obviously, the parameter dependent constraints are given in terms of currents depending on either
$\sigma$ or $-\sigma$. Therefore,
in order to obtain the constraints in terms of the independent 
canonical variables,
it is useful to 
divide the latter  into their even and odd parts, with respect to $\sigma$.
For the initial coordinates one has
\begin{equation}\label{eq:xqq}
x^\mu=q^\mu+\bar{q}^\mu,\quad
q^\mu=\sum_{n\geq0}\frac{\sigma^{2n}}{(2n)!}\, x^{(2n)\mu}
\Big{|}_{\sigma=0},
\quad
\bar{q}^\mu=
\sum_{n\geq0}\frac{\sigma^{2n+1}}{(2n+1)!}\, x^{(2n+1)\mu}
\Big{|}_{\sigma=0},
\end{equation}
and for the momenta one has
\begin{equation}\label{eq:pipp}
\pi_\mu=p_\mu+\bar{p}_\mu,\quad
p_\mu=\sum_{n\geq0}\frac{\sigma^{2n}}{(2n)!}\, \pi^{(2n)}_{\mu}
\Big{|}_{\sigma=0},
\quad
\bar{p}_\mu=
\sum_{n\geq0}\frac{\sigma^{2n+1}}{(2n+1)!}\, \pi^{(2n+1)}_{\mu}
\Big{|}_{\sigma=0}.
\end{equation}

It is well known that this separation, leads to a solvable form of the constraints which now read
\begin{eqnarray}\label{eq:gnd}
&&\Gamma^{N}_{a}(\sigma)=
2(BG^{-1})_{a}^{\ b}\,p_b
+\bar{p}_{a}
-\kappa(G_{E})_{ab}\bar{q}^{\prime b},
\nonumber\\
&&\Gamma^{i}_{D}(\sigma)=
(G^{-1})^{ij}\Big[
p_{j}-\kappa G_{jk}q^{\prime k}
+2\kappa B_{jk}\bar{q}^{\prime k}
\Big].
\end{eqnarray}
Using the above expressions 
for the constraints of the initial theory
\begin{equation}\label{eq:pock}
\Gamma^{N}_{a}(\sigma)=0,
\quad
\Gamma^{i}_{D}(\sigma)=0,
\end{equation}
one obtains the solution
\begin{eqnarray}\label{eq:sol1}
\bar{p}_{a}=0,&&\bar{q}^{\prime a}=-\theta^{ab}p_{b},
\nonumber\\
q^{\prime i}=0,&&p_{i}=-2\kappa B_{ij}\bar{q}^{\prime j}.
\end{eqnarray}
\subsubsection{The constraints at $\sigma=\pi$}

In order to derive constraints at the other string end-point $\sigma=\pi$,
we will multiply every constraint with the appropriate power of $\sigma-\pi$
and sum the products to obtain
two sigma dependent constraints
\begin{equation}
_{\pi}\Gamma^{N}_{a}(\sigma)=\sum_{n\geq0}\frac{(\sigma-\pi)^{n}}{n!}\, 
{_{_{N}}}\!\gamma^{n}_{a}
\Big{|}_{\sigma=\pi},
\qquad
_{\pi}\Gamma^{i}_{D}(\sigma)=\sum_{n\geq0}\frac{(\sigma-\pi)^{n}}{n!}\,
{_{_{D}}}\!\gamma_{n}^{i}\Big{|}_{\sigma=\pi}.
\end{equation}
Substituting the canonical form of the constraints (\ref{eq:conj}), we obtain
\begin{eqnarray}
&&_\pi\Gamma^{N}_{a}(\sigma)=\sum_{n\geq0}\frac{(\sigma-\pi)^{n}}{n!}
\Big[
\Pi_{+ab}(G^{-1})^{bc}j^{(n)}_{-c}
+\Pi_{-ab}(G^{-1})^{bc}(-1)^{n}j_{+c}^{(n)}
\Big]\Big{|}_{\sigma=\pi},
\nonumber\\
&&_\pi\Gamma^{i}_{D}(\sigma)=\frac{1}{2}\sum_{n\geq0}\frac{(\sigma-\pi)^{n}}{n!}
(G^{-1})^{ij}\Big[(-1)^{n}j_{+j}^{(n)}+j_{-j}^{(n)}\Big]\Big{|}_{\sigma=\pi}.
\end{eqnarray}
Summing, we obtain the explicit form of the sigma dependent constraints
\begin{eqnarray}\label{eq:gammapi}
&&_\pi\Gamma^{N}_{a}(\sigma)=
\Pi_{+ab}(G^{-1})^{bc}j_{-c}(\sigma)
+\Pi_{-ab}(G^{-1})^{bc}
j_{+c}(2\pi-\sigma)
\nonumber\\
&&_\pi\Gamma^{i}_{D}(\sigma)=\frac{1}{2}
(G^{-1})^{ij}
\Big[
j_{+j}(2\pi-\sigma)
+j_{-j}(\sigma)\Big].
\end{eqnarray}

Comparing the constraints (\ref{eq:gammapi}) and (\ref{eq:gamma}),
one observes that they are equal if
\begin{eqnarray}
&& j_{+a}(2\pi-\sigma)=j_{+a}(-\sigma),
\nonumber\\
&&j_{+i}(2\pi-\sigma)=j_{+i}(-\sigma).
\end{eqnarray}

It follows that if we extend the domain \cite{DS} of the
variables building the currents, i.e. original coordinates and momenta and
demand their $2\pi$-perio\-dicity
\begin{eqnarray}
&&
x^\mu(\sigma+2\pi)=x^\mu(\sigma),
\nonumber\\
&&
\pi_\mu(\sigma+2\pi)=\pi_\mu(\sigma),
\end{eqnarray}
then the $\sigma-$dependent constraints for $\sigma=0$ and $\sigma=\pi$ are equal,
and their solution is given by (\ref{eq:sol1}).

\subsection{T-dual theory constraints}

The canonical form of the T-dual boundary conditions (\ref{eq:dualbc}) is obtained
using the expression for the T-dual momenta, given by the second relation of (\ref{eq:momenta})
and the dual currents (\ref{eq:strujadual}).
The conditions are rewritten as 
\begin{equation}\label{eq:nbs}
^\star_{\mathsmaller N}\gamma^{i}_{0}=
\frac{\kappa}{2}\Big[
\Theta_{-}^{ij}
G^{E}_{jk}\,{^\star j^k_{-}}
+\Theta_{+}^{ij}
G^{E}_{jk}\,{^\star j^k_{+}}
\Big],
\end{equation}
and
\begin{equation}\label{eq:dbs}
_{\mathsmaller D}^\star\gamma_{a}^{0}=
\frac{1}{2}G^{E}_{ab}\Big{(}
{^\star j^b_{+}}
+{^\star j^b_{-}}
\Big{)}.
\end{equation}
Although the boundary conditions of the initial and the T-dual theory are related by (\ref{eq:bcrel}),
so that the Neumann and the Dirichlet conditions of the initial theory transform to the 
Dirichlet and the Neumann conditions in the T-dual theory,
one can observe that the form of Neumann and Dirichlet conditions has not changed.
Rewriting (\ref{eq:nbs}) and (\ref{eq:dbs}) as
\begin{eqnarray}
&&^\star_{\mathsmaller N}\gamma^{i}_{0}=\,^\star\Pi_{+ij}(^\star G^{-1})^{jk}\,{^\star j}_{-k}
+^\star\Pi_{-ij}(^\star G^{-1})^{jk}\,{^\star j}_{+k},
\nonumber\\
&&
_{\mathsmaller D}^\star\gamma_{a}^{0}=
\frac{1}{2}
(^\star G^{-1})_{ab}
\Big{(}
{^\star j^b_{+}}
+{^\star j^b_{-}}
\Big{)}.
\end{eqnarray}
using ${^\star\Pi}_{\pm}^{ij}=\frac{\kappa}{2}\Theta^{ij}_\mp$ and ${^\star G}^{\mu\nu}=(G^{-1}_{E})^{\mu\nu}$,
we see that they are of the same form as (\ref{eq:bccan})
keeping in mind the T-duality relations
\begin{equation}
\Pi_{\pm \mu\nu}\rightarrow {^\star\Pi}_{\pm}^{\mu\nu},
\quad
G_{\mu\nu}\rightarrow{^\star G}^{\mu\nu},
\quad
j_{\pm\mu}\rightarrow {^\star j}^\mu_\pm.
\end{equation}

Using the Dirac procedure, analogue to that for the initial theory,  the following $\sigma$-dependent constraints are
obtained
\begin{eqnarray}
^\star\Gamma_{N}^{i}(\sigma)&=&
\frac{\kappa}{2}\Big[
\Theta_{-}^{ij}
G^{E}_{jk}{^\star j^{k}_{-}(\sigma)}
+\Theta_{+}^{ij}
G^{E}_{jk}{^\star j^{k}_{+}(-\sigma)}
\Big],
\nonumber\\
^\star\Gamma_{a}^{D}(\sigma)&=&
\frac{1}{2}G^{E}_{ab}\big{[}
{^\star j^{b}_{+}}(-\sigma)
+{^\star j^{b}_{-}}(\sigma)
\big{]}.
\end{eqnarray}
The constraints at $\sigma=\pi$ produce the same result if we demand the $2\pi$-periodicity for the T-dual canonical variables
$y_\mu(\sigma+2\pi)=y_\mu(\sigma)$ and $^\star\pi^\mu(\sigma+2\pi)=\,^\star\pi^\mu(\sigma)$.

Separating the dual variables into the odd and even parts with respect to $\sigma=0$,
in a same way as in (\ref{eq:xqq}) and (\ref{eq:pipp})
\begin{eqnarray}
&&y_\mu=k_\mu+\bar{k}_\mu,
\nonumber\\
&&^\star\pi^\mu={^\star{p}}^\mu+{^\star\bar{p}}^\mu,
\end{eqnarray}
one obtains the sigma dependent constraints of the following form
\begin{eqnarray}\label{eq:DC}
^\star\Gamma_{N}^{i}(\sigma)&=&-2(G^{-1}B)^{i}_{\ j}{^\star p}^j
+{^\star\bar{p}}^{i}-\kappa(G^{-1})^{ij}\bar{k}^\prime_j,
\nonumber\\
^\star\Gamma^{D}_{a}(\sigma)&=&G^{E}_{ab}\Big{[}
{^\star p}^b+\kappa^{2}\theta^{bc}\bar{k}^\prime_c
-\kappa(G^{-1}_{E})^{bc}k^\prime_c
\Big{]}.
\end{eqnarray}
The T-dual constraints are also of the second class. So, we can
solve them
\begin{equation}\label{eq:dualk}
^\star\Gamma^{D}_{a}(\sigma)=0,
\quad
^\star\Gamma^{i}_{N}(\sigma)=0,
\end{equation}
by
\begin{eqnarray}\label{eq:sold}
&&{^\star\bar{p}}^{i}=0,\qquad\bar{k}^\prime_{i}=-\frac{2}{\kappa}B_{ij}{^\star p}^{j},
\nonumber\\
&&{^\star p}^{a}=-\kappa^{2}\theta^{ab}\bar{k}^\prime_{b},\qquad
k^{\prime}_{a}=0.
\end{eqnarray}

\section {Noncommutativity of the effective variables}\label{sec:nev}
\cleq

Solving the constraints (\ref{eq:pock}), has reduced the phase space by half.
The $\sigma$-derivative of coordinates and the momenta for the solution (\ref{eq:sol1}) are
\begin{eqnarray}\label{eq:resenjex}
x^{\prime\mu}=
\begin{cases}
q^{\prime a}-\theta^{ab}p_{b},& {\mathsmaller{\mu=a}},\\
                               &                                                        \\
\bar{q}^{\prime i},& {\mathsmaller{\mu=i}},
\end{cases}
\end{eqnarray}
and
\begin{eqnarray}\label{eq:resenjepi}
\pi_\mu=
\begin{cases}
p_{a},& {\mathsmaller{\mu=a}},\\&\\
\bar{p}_{i}-2\kappa B_{ij}\bar{q}^{\prime j},& {\mathsmaller{\mu=i}}.
\end{cases}
\end{eqnarray}

By solving the constraints,
one has eliminated 
 parts of initial coordinates $\bar{q}^{a},\,q^{i}$ and momenta
$\bar{p}_{a},\,p_{i}$,
and one is left with
variables $q^{a},\bar{q}^{i},p_{a},\bar{p}_{i}$
 which are considered as fundamental variables. 
Note that in $N$-sector the new fundamental variables are even $q^{a},p_{a}$,
while in $D$-sector the new fundamental variables are odd $\bar{q}^{i},\bar{p}_{i}$.

For an arbitrary function $F(x,\pi)$ defined
on the initial phase space, one introduces its restriction on the
reduced phase space by $f=F(x,\pi)\Big{|}_{\Gamma_\mu=0}$.
The Poisson brackets in the effective phase
space are the Dirac brackets \cite{HT} of the initial phase space
associated with the second class constraints $\Gamma_\mu=0$.
The new brackets are denoted by star
\begin{equation}\label{eq:starb}
{}^{\star}\{f,g\}=\{F,G\}_{Dirac}\Big{|}_{\Gamma_\mu=0}.
\end{equation}
The Poisson brackets of the effective variables are considered in app. \ref{sec:dod} (for details see \cite{DS}).

So, by looking at the solution of the constraints, we can observe that
in the Neumann-subspace, the coordinates
depend on both effective coordinates and momenta,
while in the Dirichlet-subspace
the momenta depend on both effective coordinates and momenta.
Therefore,
using (\ref{eq:brac})
we can conclude that in the $N-$subspace, the coordinates do not commute
\begin{equation}\label{eq:xcom}
^\star\{x^{a}(\sigma),x^{b}(\bar\sigma)\}=2\theta^{ab}\theta(\sigma+\bar\sigma),
\end{equation}
while in the $D-$subspace 
the momenta do not commute
\begin{equation}\label{eq:picom}
^\star\{\pi_{i}(\sigma),\pi_{j}(\bar\sigma)\}=4\kappa B_{ij}\delta^\prime(\sigma+\bar\sigma).
\end{equation}

The dual coordinate $\sigma$-derivative and the dual momenta for the solution (\ref{eq:sold}) of the dual constraints (\ref{eq:dualk}) are
\begin{eqnarray}\label{eq:resenjedx}
y^{\prime}_{\mu}=
\begin{cases}
\bar{k}^\prime_{a},& {\mathsmaller{\mu=a}},\\&\\
k^\prime_{i}-\frac{2}{\kappa}B_{ij}{^\star p}^{j},& {\mathsmaller{\mu=i}},
\end{cases}
\end{eqnarray}
and
\begin{eqnarray}\label{eq:resenjedpi}
^\star\pi^\mu=
\begin{cases}
{^\star\bar{p}}^{a}-\kappa^{2}\theta^{ab}\bar{k}^\prime_{b},& {\mathsmaller{\mu=a}},\\&\\
^\star{p}^{i},& {\mathsmaller{\mu=i}}.
\end{cases}
\end{eqnarray}

By solving the T-dual constraints one has eliminated the variables 
$k_{a},\bar{k}_{i},{^\star p}^{a},{^\star \bar{p}}^{i}$.
Therefore, the new fundamental variables are 
$\bar{k}_{a},{^\star \bar{p}}^{a}$ (odd) in the $D$-sector and
$k_{i},{^\star p}^{i}$(even) in the $N$-sector.

In this description, we see that the coordinates in the $D-$subspace commute while in the $N-$subspace they are not commutative
\begin{equation}\label{eq:ycom}
^\star\{y_{i}(\sigma),y_{j}(\bar\sigma)\}=\frac{4}{\kappa}B_{ij}\theta(\sigma+\bar\sigma)=
2\,{^\star \theta}_{ij}\theta(\sigma+\bar\sigma).
\end{equation}
 The momenta are commutative in the $N$-subspace, while in the $D$-subspace they are noncommutative
\begin{equation}\label{eq:piscom}
^\star\{{^\star\pi}^{a}(\sigma),{^\star\pi}^{b}(\bar\sigma)\}=2\kappa^{2}\theta^{ab}\delta^\prime(\sigma+\bar\sigma)=
4\kappa\,{^\star B}^{ab}\delta^\prime(\sigma+\bar\sigma).
\end{equation}
So, $N$ and $D$-sectors of the initial and T-dual theories replace their characteristics.
Note that in all cases the Kalb-Ramond field is the source of noncommutativity.
\section{Effective theories}\label{sec:eff}
\cleq

By extending the domain of the initial coordinates and momenta, the solution of the constraint in one string endpoint
solves the constraint in the other string endpoint as well, as in \cite{DS}.
If we substitute the solution of the constraints into the canonical hamiltonians,
we will obtain the effective hamiltonians. 
Using the equations of motion for momenta, we will find the corresponding effective lagrangians.
Effective theories describe the closed effective string.

When choosing the Neumann boundary conditions for all directions
the new basic canonical variables, the effective variables, are the even coordinate and momenta parts
$q^\mu(\sigma)$
and $p_\mu(\sigma)$.
Choosing the mixed boundary conditions 
both odd and even parts of initial coordinates and momenta remain the basic canonical variables
for some directions.
So, the effective hamiltonian for the initial theory will be given in terms of odd $\bar{q}^i,\bar{p}_i$ in Dirichlet directions and even $q^a,p_a$ in Neumann directions and the effective T-dual hamiltonian in terms of
odd $\bar{k}_{a},{^\star \bar{p}}^{a}$ in Dirichlet directions and
even $k_{i},{^\star p}^{i}$ in Neumann directions. The corresponding effective lagrangians will consequently depend on both even and odd coordinate parts $q^{a},\bar{q}^{i}$
and $k_{i},\bar{k}_{a}$.


\subsection{Effective energy-momentum tensors}

For the solution (\ref{eq:resenjex}), (\ref{eq:resenjepi}) of the boundary conditions,
 the $a-$th and the $i-$th component of the initial currents $j_{\pm\mu},$
 reduce to 
\begin{eqnarray}\label{eq:jpm}
j_{\pm a}&=&
 \mp\kappa G_{ab}\Theta^{bc}_\pm\, {j}^{\mathsmaller N}_{\pm c},
\quad
{j}^{\mathsmaller N}_{\pm c}\equiv p_{c}\pm\kappa G^{E}_{cd}q^{\prime d},
\nonumber\\
j_{\pm i}&=&\bar{p}_{i}\pm\kappa G_{ij}\bar{q}^{\prime j}
\equiv
{j}^{\mathsmaller D}_{\pm i}.
\end{eqnarray}
The energy-momentum tensor components (\ref{eq:tpm}) in a background (\ref{eq:bg}) read
\begin{equation}\label{eq:emc}
T_\pm=\mp\frac{1}{4\kappa}\Big{[}({G}^{-1})^{ab}j_{\pm  a}j_{\pm b}
+({G}^{-1})^{ij}j_{\pm i}j_{\pm j}
\Big{]},
\end{equation}
and reduce to
\begin{equation}\label{eq:teff}
T^{eff}_\pm=\mp\frac{1}{4\kappa}\Big{[}
(G^{-1}_{E})^{ab}\, j^{\mathsmaller N}_{\pm  a}j^{\mathsmaller N}_{\pm b}
+({G}^{-1})^{ij}j^{\mathsmaller D}_{\pm i}j^{\mathsmaller D}_{\pm j}
\Big{]}
\equiv T^{\mathsmaller N}_\pm+T^{\mathsmaller D}_\pm,
\end{equation}
for the solution of constraints.

The dual energy-momentum tensor components
are
\begin{eqnarray}\label{eq:emcdual}
^\star T_\pm=\mp\frac{1}{4\kappa}
\Big{[}
(^\star G^{-1})_{ab}\,{^\star j}_{\pm}^{a}\,{^\star j}_{\pm}^{b}
+
(^\star G^{-1})_{ij}\,{^\star j}_{\pm}^{i}\,{^\star j}_{\pm}^{j}
\Big{]}.
\end{eqnarray}
The dual currents reduce for the solution (\ref{eq:resenjedx}) and (\ref{eq:resenjedpi}) to
\begin{eqnarray}\label{eq:jpmdual}
&&^\star j^{a}_\pm={^\star\bar{p}}^{a}\pm\kappa (G^{-1}_{E})^{ab}\bar{k}^{\prime}_{b}\equiv
\,^\star j^{a}_{{\mathsmaller D}\pm},
\nonumber\\
&&^\star j^{i}_\pm=
\pm\kappa\Theta^{ij}_\mp G_{jk}
{^\star j}_{{\mathsmaller N}\pm}^{k},
\qquad 
{^\star j}_{{\mathsmaller N}\pm}^{k}={^\star p}^{k}\pm\kappa(G^{-1})^{kl}k^\prime_{l},
\end{eqnarray}
and therefore the energy-momentum tensor components become
\begin{eqnarray}\label{eq:teffstar}
^\star T^{eff}_\pm=\mp\frac{1}{4\kappa}
\Big{[}
(G_{E})_{ab}\,{^\star j}_{{\mathsmaller D}\pm}^{a}\,{^\star j}_{{\mathsmaller D}\pm}^{b}
+G_{ij}\,{^\star j}_{{\mathsmaller N}\pm}^{i}{^\star j}_{{\mathsmaller N}\pm}^{j}
\Big{]}
\equiv\, {^\star T}^{\mathsmaller D}_\pm+^\star T^{\mathsmaller N}_\pm.
\end{eqnarray}
Note that in opposite to the initial currents,  the T-dual currents with index $a$ are Dirichlet's, while the currents with index $i$ are Neumann's.

\subsection{Effective hamiltonians}

The effective canonical hamiltonian for theory (\ref{eq:action0}) is
\begin{equation}\label{eq:hamin}
{\cal H}^{eff}=T^{eff}_{-}-T^{eff}_{+},
\end{equation}
and the effective T-dual canonical hamiltonian for (\ref{eq:tdual}) is
\begin{equation}\label{eq:hamind}
{^\star{\cal H}}_{C}^{eff}={^\star T}^{eff}_{-}-{^\star T}^{eff}_{+}.
\end{equation}

The effective hamiltonian (\ref{eq:hamin}), expressed in terms of effective variables with the help of (\ref{eq:jpm}),
reads
\begin{equation}\label{eq:heff}
{\cal H}^{eff}={\cal H}_{\mathsmaller N}^{eff}(q^{a},p_{a})
+{\cal H}_{\mathsmaller D}^{eff}(\bar{q}^{i},\bar{p}_{i}),
\end{equation}
where
\begin{eqnarray}
&&{\cal H}_{\mathsmaller N}^{eff}(q^{a},p_{a})=\frac{\kappa}{2}\,q^{\prime a}G^{E}_{ab}q^{\prime b}+\frac{1}{2\kappa}p_{a}(G^{-1}_{E})^{ab}p_{b},
\nonumber\\
&&{\cal H}_{\mathsmaller D}^{eff}(\bar{q}^{i},\bar{p}_{i})=\frac{\kappa}{2}\,\bar{q}^{\prime i}G_{ij}\bar{q}^{\prime j}+\frac{1}{2\kappa}\bar{p}_{i}(G^{-1})^{ij}\bar{p}_{j}.
\nonumber\\
\end{eqnarray}

The effective T-dual hamiltonian (\ref{eq:hamind}), expressed in terms of effective variables with the help of (\ref{eq:jpmdual}),
reads
\begin{equation}\label{eq:heffdual}
^\star{\cal H}^{eff}={^\star{\cal H}}_{\mathsmaller D}^{eff}(\bar{k}_{a},{^\star\bar{p}}^{a})
+{^\star{\cal H}}_{\mathsmaller N}^{eff}(k_{i},{^\star p}^{i}),
\end{equation}
where
\begin{eqnarray}
&&{^\star{\cal H}}_{\mathsmaller D}^{eff}(\bar{k}_{a},{^\star\bar{p}}^{a})=
\frac{\kappa}{2}\,\bar{k}^{\prime}_{ a}(G_{E}^{-1})^{ab}\bar{k}^{\prime}_{ b}
+\frac{1}{2\kappa}{^\star\bar{p}}^{a}(G_{E})_{ab}{^\star\bar{p}}^{b},
\nonumber\\
&&{^\star{\cal H}}_{\mathsmaller N}^{eff}(k_{i},{^\star p}^{i})=
\frac{\kappa}{2}\,k^{\prime}_{ i}(G^{-1})^{ij}k^{\prime}_{ j}+\frac{1}{2\kappa}{^\star p}^{i}G_{ij}{^\star p}^{j}.
\nonumber\\
\end{eqnarray}

\subsection{Effective Lagrangians }

The lagrangians of the effective theories (\ref{eq:heff}) and (\ref{eq:heffdual})  are given by
\begin{equation}\label{eq:leff}
{\cal{L}}^{eff}=\Big[\pi_\mu{\dot{x}}^\mu-{\cal{H}}_{c}(x,\pi)\Big]
\Big{|}_{\Gamma_{\mu}=0},
\end{equation}
\begin{equation}\label{eq:leffdual}
^\star{\cal{L}}^{eff}=\Big[\,{^\star\pi}^\mu{\dot{y}}_\mu-{^\star\cal{H}}_{c}(y,{^\star\pi})\Big]
\Big{|}_{^\star\Gamma_{\mu}=0}.
\end{equation}
The effective lagrangians can be separated into
\begin{eqnarray}
&&{\cal{L}}^{eff}={\cal L}_{\mathsmaller N}(q,p)+{\cal L}_{\mathsmaller D}(\bar{q},\bar{p}),
\nonumber\\
&&^\star{\cal{L}}^{eff}=
{^\star \cal L}_{\mathsmaller D}(\bar{k},{^\star \bar{p}})
+{^\star \cal L}_{\mathsmaller N}(k,{^\star p}),
\end{eqnarray}
with
\begin{eqnarray}\label{eq:lagND}
{\cal L}_{\mathsmaller N}(q,p)=p_{a}\dot{q}^{a}-{\cal H}^{eff}_{\mathsmaller N}(q^{a},p_{a}),
&&{\cal L}_{\mathsmaller D}(\bar{q},\bar{p})=\bar{p}_{i}\dot{\bar{q}}^{i}-{\cal H}^{eff}_{\mathsmaller D}(\bar{q}^{i},\bar{p}_{i}),
\nonumber\\
{^\star \cal L}_{\mathsmaller D}(\bar{k},{^\star\bar{p}})=
{^\star\bar{p}}^{a}\dot{\bar{k}}_{a}
-{^\star\cal H}^{eff}_{\mathsmaller D}(\bar{k}_{a},{^\star\bar{p}}^{a}),
&&{^\star \cal L}_{\mathsmaller N}(k,{^\star p})=
{^\star p}_{i}\dot{k}_{i}-{^\star \cal H}^{eff}_{\mathsmaller N}(k_{i},{^\star p}^{i}).
\end{eqnarray}

The explicit forms of the effective lagrangians are found by eliminating the momenta from
(\ref{eq:lagND}),
using the equations of motion for them
\begin{equation}\label{eq:eqpi}
p_{a}=\kappa G^{E}_{ab}\dot{q}^{b},
\quad
\bar{p}_{i}=\kappa G_{ij}\dot{\bar{q}}^{j},
\end{equation}
and
\begin{equation}\label{eq:eqpidual}
^\star\bar{p}^{a}=\kappa(G^{-1}_{E})^{ab}\dot{\bar{k}}_{b},
\quad
^\star p^{i}=\kappa(G^{-1})^{ij}\dot{k}_{j}.
\end{equation}
For these equations the $\sigma$-derivatives of the initial and T-dual coordinates,
given by (\ref{eq:resenjex}) and (\ref{eq:resenjedx}), 
become
\begin{eqnarray}\label{eq:xp}
x^{\prime\mu}=
\begin{cases}
q^{\prime a}+
2(G^{-1}B)^{a}_{\ b}\,
\dot{q}^{b},& {\mathsmaller{\mu=a}},\\
                               &          \\
\bar{q}^{\prime i},& {\mathsmaller{\mu=i}},
\end{cases}
\end{eqnarray}
and
\begin{eqnarray}\label{eq:res}
y^{\prime}_{\mu}=
\begin{cases}
\bar{k}^\prime_{a},& {\mathsmaller{\mu=a}},\\&\\
k^\prime_{i}-2(BG^{-1})_{i}^{\ j}
\dot{k}_{j},& {\mathsmaller{\mu=i}}.
\end{cases}
\end{eqnarray}
In order to find the expression for the initial and the T-dual coordinate
we need to introduce a double coordinate $\tilde{q}^{a}$ of the even part  of the initial coordinate $q^{a}$ 
\begin{eqnarray}
\dot{\tilde{q}}^{\,a}=q^{\prime a},\quad \tilde{q}^{\,\prime a}=\dot{q}^{a},
\end{eqnarray}
and a double coordinate $\tilde{k}_{i}$ of the even part  of the T-dual coordinate $k_{i}$
\begin{eqnarray}\label{eq:dual}
\dot{\tilde{k}}_{\,i}=k^{\prime}_{i},\quad \tilde{k}^{\,\prime}_{i}=\dot{k}_{i}.
\end{eqnarray}
The coordinates become
\begin{eqnarray}\label{eq:resx}
x^{\mu}=
\begin{cases}
q^{a}+
2(G^{-1}B)^{a}_{\ b}\,
\tilde{q}^{b},& {\mathsmaller{\mu=a}},\\
                               &                           \\
\bar{q}^{ i},& {\mathsmaller{\mu=i}},
\end{cases}
\end{eqnarray}
and
\begin{eqnarray}\label{eq:resdual}
y_{\mu}=
\begin{cases}
\bar{k}_{a},& {\mathsmaller{\mu=a}},\\&\\
k_{i}-2(BG^{-1})_{i}^{\ j}\,
\tilde{k}_{j},& {\mathsmaller{\mu=i}}.
\end{cases}
\end{eqnarray}

For the equations (\ref{eq:eqpi}) and (\ref{eq:eqpidual}),
the currents (\ref{eq:jpm}) and (\ref{eq:jpmdual}) reduce to
\begin{eqnarray}
{j}^{\mathsmaller N}_{\pm a}=\kappa(G_{E})_{ab}\,\partial_\pm q^b,
&&
j^{\mathsmaller D}_{\pm i}=\kappa G_{ij}\,\partial_\pm \bar{q}^j,
\end{eqnarray}
and
\begin{eqnarray}
^\star j^{a}_{{\mathsmaller D}\pm}=\kappa(G^{-1}_{E})^{ab}\,\partial_\pm \bar{k}_{b},
&&
{^\star j}_{{\mathsmaller N}\pm}^{i}=\kappa(G^{-1})^{ij}\,\partial_\pm k_{j}.
\end{eqnarray}

So, after elimination of the momenta  the effective lagrangians become
\begin{eqnarray}\label{eq:lageff}
&&{\cal{L}}^{eff}={\cal L}_{\mathsmaller N}(q)+{\cal L}_{\mathsmaller D}(\bar{q}),
\nonumber\\
&&^\star{\cal{L}}^{eff}=
{^\star \cal L}_{\mathsmaller D}(\bar{k})
+{^\star \cal L}_{\mathsmaller N}(k),
\end{eqnarray}
where the lagrangians (\ref{eq:lagND}) reduced to
\begin{eqnarray}
{\cal L}_{\mathsmaller N}(q)=
\frac{\kappa}{2}G^{E}_{ab}\,\eta^{\alpha\beta}
\partial_{\alpha}q^{a}\partial_{\beta}q^{b},
&&{\cal L}_{\mathsmaller D}(\bar{q})=
\frac{\kappa}{2}G_{ij}\,\eta^{\alpha\beta}
\partial_{\alpha}\bar{q}^{i}\partial_{\beta}\bar{q}^{j},
\nonumber\\
{^\star \cal L}_{\mathsmaller D}(\bar{k})=
\frac{\kappa}{2}(G^{-1}_{E})^{ab}\,\eta^{\alpha\beta}\partial_{\alpha}{\bar{k}}_{a}\partial_{\beta}{\bar{k}}_{b},
&&{^\star \cal L}_{\mathsmaller N}(k)=
\frac{\kappa}{2}(G^{-1})^{ij}\,\eta^{\alpha\beta}
\partial_{\alpha}k_{i}\partial_{\beta}k_{j}.
\end{eqnarray}

\subsection{T-duality between effective theories}

Let us now introduce coordinates
\begin{align}\label{eq:coordvec}
    Q^\mu = \begin{bmatrix}
           q^a \\
           \bar{q}^i
         \end{bmatrix},
&\quad
K_\mu = \begin{bmatrix}
           {\bar{k}}_{a} \\
           k_{i}
         \end{bmatrix},
  \end{align}
and the corresponding canonically conjugated momenta
\begin{align}\label{eq:pivec}
    P_\mu = \begin{bmatrix}
           p_a \\
           \bar{p}_i
         \end{bmatrix},
&\quad
^\star P^\mu = \begin{bmatrix}
           ^\star{\bar{p}}^{a} \\
           ^\star{p}^{i}
         \end{bmatrix}.
  \end{align}
The currents ${j}^{\mathsmaller N}_{\pm a}$ and ${j}^{\mathsmaller D}_{\pm i}$
defined in  (\ref{eq:jpm}) and the currents $^\star j^{a}_{{\mathsmaller D}\pm}$ and $^\star j^{i}_{{\mathsmaller n}\pm}$ defined in  (\ref{eq:jpmdual}), can be gathered into currents
 \begin{align}\label{eq:coordvec}
    \hat{j}_{\pm\mu} = \begin{bmatrix}
           {j}^{\mathsmaller N}_{\pm a} \\
           {j}^{\mathsmaller D}_{\pm i}
         \end{bmatrix},
&\quad
^\star\hat{j}^\mu_\pm = \begin{bmatrix}
           ^\star j^{a}_{{\mathsmaller D}\pm} \\
           ^\star j^{i}_{{\mathsmaller N}\pm}
         \end{bmatrix}.
  \end{align}
They satisfy
\begin{eqnarray}
&&\hat{j}_{\pm\mu} = P_\mu\pm\kappa G^{eff}_{\mu\nu}Q^{\prime\nu},
\nonumber\\
&&^\star\hat{j}^\mu_\pm =\,^\star P^\mu\pm\kappa\, {^\star G}_{eff}^{\mu\nu}K^\prime_\nu,
\end{eqnarray}
where
\begin{eqnarray}\label{eq:matriceseff}
G^{eff}_{\mu\nu}=\left(\begin{array}{cc}
G^{E}_{ab} & 0\\
0 & G_{ij}
\end{array}\right),
&&
{^\star G}_{eff}^{\mu\nu}=
\left(\begin{array}{cc}
(G^{-1}_{E})^{ab} & 0\\
0 & (G^{-1})^{ij}
\end{array}\right).
\end{eqnarray}
The effective energy-momentum components (\ref{eq:teff}) and (\ref{eq:teffstar}) can be rewritten as
\begin{eqnarray}
&&T^{eff}_\pm=\mp\frac{1}{4\kappa}
(G_{eff}^{-1})^{\mu\nu}\,\hat{j}_{\pm\mu}\,\hat{j}_{\pm\nu}\,,
\nonumber\\
&&^\star T^{eff}_\pm=\mp\frac{1}{4\kappa}
(^\star G_{eff}^{-1})_{\mu\nu}\,^\star\hat{j}_{\pm}^{\mu}\,^\star\hat{j}_{\pm}^{\nu}\,,
\end{eqnarray}
and the effective hamiltonians (\ref{eq:heff}) and (\ref{eq:heffdual}) are therefore
\begin{eqnarray}\label{eq:hek}
&&{\cal H}^{eff}=\frac{\kappa}{2}Q^{\prime\mu}G^{eff}_{\mu\nu}Q^{\prime\nu}
+\frac{1}{2\kappa}P_\mu(G^{-1}_{eff})^{\mu\nu}P_\nu,
\nonumber\\
&&^\star {\cal H}^{eff}=\frac{\kappa}{2}K^{\prime}_\mu {^\star G}_{eff}^{\mu\nu}K^{\prime}_{\nu}
+\frac{1}{2\kappa}{^\star P}^\mu(^\star G^{-1}_{eff})_{\mu\nu}{^\star P}^\nu.
\end{eqnarray}

Using the T-duality
relations
\begin{equation}
\kappa x^{\prime\mu}\cong\,^\star\pi^\mu,
\quad
\kappa y^{\prime}_{\mu}\cong\pi_\mu,
\end{equation}
and (\ref{eq:resenjex}), (\ref{eq:resenjedpi}), (\ref{eq:resenjedx}), (\ref{eq:resenjepi})  one obtains
\begin{eqnarray}
&&\kappa q^{\prime a}-\kappa\theta^{ab}p_{b}\cong {^\star\bar{p}}^{a}-\kappa^{2}\theta^{ab}\bar{k}^\prime_{b},
\nonumber\\
&&\kappa \bar{q}^{\prime i}\cong\,^\star{p}^{i},
\end{eqnarray}
and
\begin{eqnarray}
&&\kappa \bar{k}^\prime_{a}\cong p_{a},
\nonumber\\
&&\kappa k^\prime_{i}-2B_{ij}{^\star p}^{j}\cong \bar{p}_{i}-2\kappa B_{ij}\bar{q}^{\prime j}.
\end{eqnarray}
Separating the odd and even parts one obtains
\begin{eqnarray}
&&\kappa q^{\prime a}\cong {^\star\bar{p}}^{a},
\quad
\kappa\bar{k}^\prime_{a}\cong p_{a},
\nonumber\\
&&\kappa \bar{q}^{\prime i}\cong\,^\star{p}^{i},
\quad
\kappa k^\prime_{i}\cong\bar{p}_{i},
\end{eqnarray}
which gives 
\begin{equation}\label{eq:tlaw}
\kappa Q^{\prime\mu}\cong\,^\star P^\mu,
\quad \kappa K^{\prime}_{\mu}\cong\,P_\mu.
\end{equation}
Comparing the background fields (\ref{eq:matriceseff}), we see that they are T-dual to each other as expected,
because by T-duality the metric should transform to the inverse of the effective metric.
In our case, in absence of the effective Kalb-Ramond field this means the T-dual metric  should be inverse to the initial metric,
what is just the case
\begin{eqnarray}\label{eq:tlawG}
(G^{eff}_{\mu\nu})^{-1}=\left(\begin{array}{cc}
G^{E}_{ab} & 0\\
0 & G_{ij}
\end{array}\right)^{-1}
=\left(\begin{array}{cc}
(G^{-1}_{E})^{ab} & 0\\
0 & (G^{-1})^{ij}
\end{array}\right)={^\star G}_{eff}^{\mu\nu}.
\end{eqnarray}
Using (\ref{eq:tlaw}) and (\ref{eq:tlawG}) we can conclude that the effective hamiltonians (\ref{eq:hek}) are T-dual to each other.

The corresponding lagrangians (\ref{eq:lageff}) are given by
\begin{eqnarray}
&&{\cal L}^{eff}=\dot{Q}^\mu P_\mu-{\cal H}^{eff}(Q,P),
\nonumber\\
&&{^\star{\cal L}}^{eff}=\dot{K}_\mu{^\star P}^\mu-\,^\star{\cal H}^{eff}(K,\,^\star P),
\end{eqnarray}
which for the equations of motion for momenta (\ref{eq:eqpi}) and (\ref{eq:eqpidual})
\begin{equation}\label{eq:eqP}
P_\mu=\kappa G^{eff}_{\mu\nu}\dot{Q}^\nu,\quad
^\star P_\mu=\kappa\,^\star G^{eff}_{\mu\nu}\dot{K}_\nu,
\end{equation}
become
\begin{eqnarray}\label{eq:eflk}
&&{\cal L}^{eff}=
\frac{\kappa}{2}\,\eta^{\alpha\beta}\,
\partial_\alpha Q^\mu\,
G^{eff}_{\mu\nu}\,
\partial_\beta Q^\nu,
\nonumber\\
&&{^\star{\cal L}}^{eff}=
\frac{\kappa}{2}\,\eta^{\alpha\beta}\,
\partial_\alpha K_\mu\,
{^\star G}_{eff}^{\mu\nu}\,
\partial_\beta K_\nu.
\end{eqnarray}
Combining (\ref{eq:tlaw}) with (\ref{eq:eqP}) one obtains
\begin{equation}
Q^{\prime\mu}\cong{^\star G_{eff}^{\mu\nu}}\dot{K}_\nu,
\quad
K^{\prime}_{\mu}\cong G^{eff}_{\mu\nu}\dot{Q}^\nu.
\end{equation}
Therefore, the effective and T-dual effective  variables $Q^\mu$ and $K_\mu$ are connected by 
\begin{equation}\label{eq:QKt}
\partial_\pm K_\mu\cong\pm G^{eff}_{\mu\nu}\partial_\pm Q^\nu.
\end{equation}
This is the T-dual effective coordinate transformation  law. Using it together with (\ref{eq:tlawG}), one can conclude that the effective lagrangians (\ref{eq:eflk}) are T-dual.
This law is in agreement with the T-dual coordinate transformation law (\ref{eq:rel}),
for $B_{\mu\nu}=0$
\begin{eqnarray}
\partial_{\pm}y_\mu\cong
\pm G_{\mu\nu}\partial_{\pm}x^\nu,
\end{eqnarray}
keeping in mind that the metric is replaced by the effective metric $G_{\mu\nu}\rightarrow G^{eff}_{\mu\nu}$.

\section{Conclusion}
\cleq

In the present paper we show that 
solving the constraints obtained applying the Dirac consistency procedure to mixed boundary conditions of the open bosonic string, which leads to the effective theory and the T-dualization of the bosonic string theory can be performed in an arbitrary order.
We started considering 
the string described by
the open string sigma model.
The string is moving in the constant metric $G_{\mu\nu}$ and a constant Kalb-Ramond field $B_{\mu\nu}$.
We chose the Neumann boundary conditions for some directions $x^{a}$ and the
Dirichlet boundary conditions for all other directions $x^{i}$.

We treated the boundary conditions as constraints, and applied the Dirac procedure.
The boundary conditions  where given in terms of coordinates and momenta, which we rewrote 
in terms of currents building the energy-momentum tensor components.
By Dirac procedure
the new constraints are found commuting the hamiltonian with the known constraints.
The canonical form of constraints 
allowed us a simple calculation of the exact form of the infinitely many constraints.
From these constraints we formed two $\sigma$-dependent constraints, for every string endpoint,
by multiplying every obtained constraint with the appropriate power of $\sigma$ for constraints in $\sigma=0$ and $\pi-\sigma$ for constraints in $\sigma=\pi$, and
adding these terms together into Taylor expansions.
The constraints at $\sigma=0$ and $\sigma=\pi$ were found to be equivalent by imposing $2\pi$-periodicity condition for the canonical variables
$x^\mu$ and $\pi_\mu$.

The $\sigma$-dependent  constraints are of the second class.
To solve them we introduced even and odd parts of the initial canonical variables.
We found the solution and expressed
the $\sigma$-derivative of the initial coordinate $x^{\prime\mu}$ and the initial momentum $\pi_\mu$  in terms
of  even parts $q^{a},p_{a}$ of $x^\mu,\pi_\mu$  in Neumann directions
and of their odd parts $\bar{q}^{i},\bar{p}_{i}$ in Dirichlet directions,
see (\ref{eq:resenjex}) and (\ref{eq:resenjepi}.)
For the solution of constraints,
the theory reduced to the effective theory.
We obtained the effective energy-momentum tensors (\ref{eq:teff}) and the effective hamiltonian (\ref{eq:heff}).
For the equations of motion for momenta, we obtained the corresponding effective lagrangian (\ref{eq:lageff}).

We also found the T-dual of the initial theory.
We applied the Dirac procedure to the mixed boundary conditions of the T-dual theory. 
The constraints where solved, which reduced the phase space to $\bar{k}_{a},{^\star \bar{p}}^{a}$ in
$D$-sector and $k_{i},{^\star p}^{i}$ in $N$-sector.
For the solution of 
T-dual constraints
 we obtained the T-dual effective energy-momentum tensors (\ref{eq:teffstar}) and the T-dual effective hamiltonian (\ref{eq:heffdual}),
as well as the corresponding T-dual effective lagrangian (\ref{eq:lageff}).

The canonically conjugated effective variables are now pairs $q^{a},p_{a}$ and $\bar{q}^{i},\bar{p}_{i}$
for the initial and
$k_{i},{^\star p}^{i}$ and $\bar{k}_{a},{^\star \bar{p}}^{a}$
 for the T-dual effective theory.
The effective variables in both effective theories satisfy the modified Poisson brackets considered in appendix \ref{sec:dod}.
Therefore, 
if the variable of the initial theory depends on both effective coordinates and effective momenta of any pair,
it will be  noncommutative.
One observes that 
in $N$-sector coordinates do not commute (\ref{eq:xcom}) and also the momenta of the $D$-sector of the initial theory (\ref{eq:picom}).
In T-dual theory the roles are exchanged so that in $D$-sector coordinates do not commute (\ref{eq:ycom}) and also
the momenta of the $N$-sector (\ref{eq:piscom}).

This is different, in comparison to the choice of the Neumann boundary conditions for all directions \cite{DS,DS1,DS2}.
In that case, solving the constraints leads to full elimination of odd variables.
Also, when considering a weakly curved background, with a coordinate dependent Kalb-Ramond field with an infinitesimal 
field strength, the 
effective theory turned out to be non-geometric. It is defined in the effective space-time composed of the even coordinate and its double
$
x^\mu\rightarrow q^\mu,\, {\tilde{q}}^\mu.
$
This fact lead to appearance of nontrivial effective Kalb-Ramond field, depending on the double effective coordinate
$
B_{\mu\nu}(x)\rightarrow B^{eff}_{\mu\nu}(2b{\tilde{q}}).$
It would be interesting to find the corresponding field in the mixed boundary conditions case.
For constant initial background fields, considered in this paper
the effective fields are constant. But, the nongeometricity can still be seen.
It appears  in a fact that coordinates of the initial and T-dual theories, can not be 
expressed without an introduction of double coordinates, see (\ref{eq:resx}) and (\ref{eq:resdual}).

The obtained effective theories, defined in terms of the effective variables, where compared using the T-dualization procedure.
It was confirmed that
the corresponding background fields (the effective metrics $G^{eff}_{\mu\nu}$ and $^\star G_{eff}^{\mu\nu}$ (\ref{eq:matriceseff})) are T-dual to each other.
Also, the effective variables of the initial effective theory  are confirmed to be T-dual to the T-dual effective variables
of the T-dual effective theory,
by obtaining the T-duality law connecting them.
This law was an appropriate reduction of the standard T-duality coordinate transformation law.
Therefore, we showed the T-duality of the reduced bosonic string theories.
Consequently, all the theories on the following diagram are equivalent

\begin{tikzpicture}
\draw (0,0) node {$\kappa\int d\xi^{2}\partial_{+}x^\mu\Pi_{+\mu\nu}\partial_{-}x^\nu$};
\draw (6,0) node {$\frac{\kappa^{2}}{2}\int d\xi^{2}\partial_{+}y_\mu
\Theta_{-}^{\mu\nu}\partial_{-}y_\nu$};
\draw (0,-2) node {$\frac{\kappa}{2}\int d\xi^{2}\partial_{+}Q^\mu G^{eff}_{\mu\nu}\partial_{-}Q^\nu$};
\draw (6,-2) node {$\frac{\kappa}{2}\int d\xi^{2}\partial_{+}K_\mu{^\star G}_{eff}^{\mu\nu}\partial_{-}K_\nu$};
\draw[thick,<->] (2.5,0) -- (3.5,0);
\draw (3,0.5) node {$T$};
\draw[thick,<->] (2.5,-2) -- (3.5,-2);
\draw (3,-1.5) node {$T$};
\draw[thick,->] (0,-0.5) -- (0,-1.5);
\draw[thick,->] (6,-0.5) -- (6,-1.5);
\draw (0.5,-1) node {$^{\Gamma=0}$};
\draw (6.5,-1) node {$^{^\star\Gamma=0}$};
\end{tikzpicture}\\

So, we confirmed that two procedures, the T-dualization procedure and 
the solving of the mixed boundary conditions, treated as constraints in the Dirac consistency procedure, do commute.

\appendix
\section{ Brackets between effective variables}\label{sec:dod}
\cleq

The effective theory is given in terms of the odd and even parts of the initial coordinates and momenta.
These parts do not satisfy the ordinary Poisson brackets because they are not the arbitrary functions,
but contain only even or odd powers of $\sigma$.
Additionally their domain is changed in order to solve the boundary conditions in both string endpoints.
The new fundamental variables satisfy
the modified Poisson brackets, defined with the appropriate delta functions.

The standard Poisson brackets between the initial coordinates and the momenta
\begin{equation}\label{eq:spb}
\{x^\mu(\sigma),\pi_\nu(\bar{\sigma})\}=\delta^\mu_\nu\delta(\sigma-\bar\sigma),
\end{equation}
give
\begin{eqnarray}\label{eq:poisson}
\{q^\mu(\sigma),p_\nu(\bar{\sigma})\}=\delta^{\mu}_{\ \nu}\delta_{S}(\sigma,\bar{\sigma}),&&
\{{\bar{q}}^\mu(\sigma),{\bar{p}}_\nu(\bar{\sigma})\}=\delta^{\mu}_{\ \nu}\delta_{A}(\sigma,\bar{\sigma}),
\end{eqnarray}
where $\delta_{S}$ and $\delta_{A}$ are defined by
\begin{eqnarray}\label{eq:dsa}
\delta_{S}(\sigma,\bar{\sigma})=
\frac{1}{2}[\delta(\sigma-\bar{\sigma})+\delta(\sigma+\bar{\sigma})],&&
\delta_{A}(\sigma,\bar{\sigma})=
\frac{1}{2}[\delta(\sigma-\bar{\sigma})-\delta(\sigma+\bar{\sigma})],
\end{eqnarray}
and the domain is $[-\pi,\pi]$.
The even and odd coordinate parts satisfy
\begin{equation}
\int_{-\pi}^{\pi} d \bar\sigma q^\mu(\bar\sigma)
\delta_{S}(\bar{\sigma},\sigma)= q^\mu(\sigma),\qquad
\int_{-\pi}^{\pi} d \bar\sigma {\bar{q}}^\mu(\bar\sigma)
\delta_{S}(\bar{\sigma},\sigma)= {\bar{q}}^\mu(\sigma).
\end{equation}

Separating integration domain in two parts, from $-\pi$ to $0$ and
from $0$ to $\pi$, and changing the integration variable in the first
part $\bar{\sigma} \to - \bar{\sigma}$, we obtain
\begin{equation}
2\int_0^{\pi} d \bar\sigma q^\mu(\bar\sigma)
\delta_{S}(\bar{\sigma},\sigma)
= q^\mu(\sigma),\qquad
2\int_0^{\pi} d \bar\sigma {\bar{q}}^\mu(\bar\sigma)
\delta_{S}(\bar{\sigma},\sigma)
= {\bar{q}}^\mu(\sigma).
\end{equation}

So, the unit functions on the interval $[0,\pi]$ for
functions with only an even or odd power in $\sigma$ are
$2 \delta_{S}(\bar{\sigma},\sigma)$
and
$2 \delta_{A}(\bar{\sigma},\sigma)$, respectively.
Therefore, the brackets
which we use are
\begin{equation}\label{eq:brac}
^\star\{q^\mu(\sigma),p_\nu(\bar{\sigma})\}= 2
\delta^{\mu}_{\nu} \delta_{S}(\sigma,\bar{\sigma}),\quad
^\star\{{\bar{q}}^\mu(\sigma),{\bar{p}}_\nu(\bar{\sigma})\}= 2
\delta^{\mu}_{\nu} \delta_{A}(\sigma,\bar{\sigma}),
\quad \sigma,
\bar{\sigma} \in[0,\pi].
\end{equation}



\begin{thebibliography}{}


\bibitem{BGS} L. Brink, M.B. Green, J.H. Schwarz, {\it Nucl. Phys.} {\bf B 198} (1982)  474.
\bibitem{KY} K. Kikkawa, M. Yamasaki, {\it Phys. Lett.} {\bf B 149} (1984)  357.
\bibitem{SS} N. Sakai, I. Senda, {\it Prog. Theor. Phys.} {\bf 75} (1984) 692.
\bibitem{B1} T. Buscher, {\it Phys. Lett.} {\bf B 194} (1987) 51.
\bibitem{B2}  T. Buscher, {\it Phys. Lett.} {\bf 201} (1988) 466.
\bibitem{RV} M. Ro\v cek, E. Verlinde,  {\it Nucl. Phys.} {\bf B 373} (1992) 630.
\bibitem{DST} Lj. Davidovi\' c and B. Sazdovi\' c, {\it Eur. Phys. J.} {\bf C 74} (2014) 2683. 
\bibitem{DFT1} O. Hohm, C. Hull and B. Zwiebach, {\it JHEP} \textbf{08} (2010) 008.
\bibitem{DFT2} O. Hohm, C. Hull and B. Zwiebach, {\it JHEP} \textbf{07} (2010) 016.
\bibitem{DS} Lj. Davidovi\' c and B. Sazdovi\' c, {\it Phys. Rev.} {\bf D 83} (2011) 066014.
\bibitem{DS1} Lj. Davidovi\' c and B. Sazdovi\' c, {\it JHEP} {\bf 08} (2011) 112.
\bibitem{DS2} Lj. Davidovi\' c and B. Sazdovi\' c, {\it Eur. Phys. J.} {\bf C 72} (2012) 2199.
\bibitem{SW} N. Seiberg and E. Witten, {\it JHEP} \textbf{09} (1999) 032.
\bibitem{SN} B. Sazdovi\'c, {\it Eur. Phys. J.} \textbf{C44} (2005) 599.
\bibitem{SN1} B. Nikoli\'c and B. Sazdovi\'c, {\it Phys. Rev.} \textbf{D74} (2006) 045024.
\bibitem{SN2} B. Nikoli\'c and B. Sazdovi\'c, {\it Phys. Rev.} \textbf{D75} (2007) 085011.
\bibitem{SN3} B. Nikoli\'c and B. Sazdovi\'c, {\it Adv. Theor. Math. Phys.} \textbf{14} (2010) 1.
\bibitem{BBS} K. Backer, M. Backer and J. Schwarz, \textit{String Theory and M-theory}, (Cambridge University Press, Cambridge, 2007);
C.V. Johnson, \textit{D-branes}, (Cambridge University Press, Cambridge, 2003).
\bibitem{Z} B. Zwiebach, {\it A First Course in String Theory} (Cambridge University Press, Cambridge, 2009).
\bibitem{S} B. Sazdovi\' c, {\it Chinese Physics} {\bf C 42} (2018) 083106.
\bibitem{DS18} Lj. Davidovi\' c and B. Sazdovi\' c, {\it Eur. Phys. J.} {\bf C 78} (2018) 600.
\bibitem{HT} M. Henneaux and C. Teitelboim, {\it Quantization of Gauge Systems}, (Princeton University Press, Princeton,1992).


\end{thebibliography}
\end{document}